\documentclass[twocolumn, tighten]{aastex631}

\graphicspath{{./}{figs/}}
\hypersetup{urlcolor=blue}

\usepackage{tabularx}

\begin{document}

\title{The far side of the Galactic bar/bulge revealed through semi-regular variables}

\author[0000-0003-3244-5357]{Daniel R. Hey}
\affil{Institute for Astronomy, University of Hawai`i, Honolulu, HI 96822, USA}

\author[0000-0003-3244-5357]{Daniel Huber}
\affiliation{Institute for Astronomy, University of Hawai`i, Honolulu, HI 96822, USA}
\affiliation{Sydney Institute for Astronomy, School of Physics, University of Sydney NSW 2006, Australia}

\author[0000-0003-3244-5357]{Benjamin J. Shappee}
\affiliation{Institute for Astronomy, University of Hawai`i, Honolulu, HI 96822, USA}

\author[0000-0001-7516-4016]{Joss Bland-Hawthorn}
\affiliation{Sydney Institute for Astronomy, School of Physics, University of Sydney NSW 2006, Australia}
\affiliation{Center of Excellence for All Sky Astrophysics in Three Dimensions (ASTRO-3D), Australia}

\author[0000-0002-1081-883X]{Thor Tepper-García}
\affiliation{Sydney Institute for Astronomy, School of Physics, University of Sydney NSW 2006, Australia}
\affiliation{Center of Excellence for All Sky Astrophysics in Three Dimensions (ASTRO-3D), Australia}

\author[0000-0003-3939-3297]{Robyn Sanderson}
\affiliation{Department of Physics \& Astronomy, University of Pennsylvania, 209 S 33rd St., Philadelphia, PA 19104, USA}
\affiliation{Center for Computational Astrophysics, Flatiron Institute, 162 5th Ave., New York, NY 10010, USA}

\author[0000-0001-6711-8140]{Sukanya Chakrabarti}
\affiliation{Department of Physics and Astronomy, University of Alabama in Huntsville}

\author[0000-0003-2657-3889]{Nicholas Saunders}
\affiliation{Institute for Astronomy, University of Hawai`i, Honolulu, HI 96822, USA}

\author[0000-0001-8917-1532]{Jason A. S. Hunt}
\affiliation{Center for Computational Astrophysics, Flatiron Institute, 162 5th Ave., New York, NY 10010, USA}

\author[0000-0001-5222-4661]{Timothy R. Bedding}
\affiliation{Sydney Institute for Astronomy, School of Physics, University of Sydney NSW 2006, Australia}

\author[0000-0003-2858-9657]{John Tonry}
\affiliation{Institute for Astronomy, University of Hawai`i, Honolulu, HI 96822, USA}


\newcommand{\kepler}{{\em Kepler\/}}
\newcommand{\gaia}{{\em Gaia\/}}
\renewcommand{\gaia}{Gaia}
\newcommand{\tess}{{\em TESS\/}}
\newcommand{\trb}[1]{{\color{magenta}\textbf{[Tim: #1]}}}

\newcommand{\nstars}{\mbox{$232,136$}}
\newcommand{\ngood}{\mbox{$190,302$}}

\newcommand{\nkinematic}{\mbox{$39,566$}}
\newcommand{\rnought}{\mbox{$8108\pm106_{\rm stat}\pm93_{\rm sys}$}}

\newcommand{\todo}[1]{{\color{red}{[TODO: #1]}}}
\newcommand{\addcite}[1]{{\color{red}{[CITE: #1]}}}

\newcommand{\ramses}{{\sc ramses}}
\newcommand{\agama}{{\sc agama}}

\newcommand{\rev}[1]{\textcolor{black}{#1}}

\begin{abstract}
    The Galactic bulge is critical to our understanding of the Milky Way. However, due to the lack of reliable stellar distances, the structure and kinematics of the bulge/bar beyond the Galactic center have remained largely unexplored. Here, we present a method to measure distances of luminous red giants using a period-amplitude-luminosity relation anchored to the Large Magellanic Cloud, with random uncertainties of 10--15\% and systematic errors below 1--2\%. We apply this method to data from the Optical Gravitational Lensing Experiment (OGLE) to measure distances to \ngood\ stars in the Galactic bulge and beyond out to 20\,kpc. Using this sample we measure a distance to the Galactic center of $R_0$ = \rnought\ pc, consistent with direct measurements of stars orbiting Sgr A*. We cross-match our distance catalog with Gaia DR3 and use the subset of \nkinematic\ overlapping stars to provide the first constraints on the Milky Way's velocity field ($V_R,V_\phi,V_z$) beyond the Galactic center. We show that the $V_R$ quadrupole from the bar's near side is reflected with respect to the Galactic center, indicating that the bar is bi-symmetric and aligned with the inner disk. We also find that the vertical height $V_Z$ map has no major structure in the region of the Galactic bulge, which is inconsistent with a current episode of bar buckling. Finally, we demonstrate with N-body simulations that distance uncertainty plays a factor in the alignment of the major and kinematic axes of the bar, necessitating caution when interpreting results for distant stars.

\end{abstract}


\section{Introduction} \label{sec:intro}

Galactic disks are cold, rotating, low-entropy systems that are highly unstable to bar instabilities \citep{hohl1971}. These instabilities grow exponentially and the timescale for the bar to emerge depends on the disk mass as a fraction of the total mass, $f_{\rm disk}$, across the inner disk \citep{fuj18a}. If the Milky Way has sustained its current high value of $f_{\rm disk}$ for most of its life \citep{Bland-Hawthorn2016Galaxy}, the bar/bulge formed early in the Universe's history and is, therefore, likely to be very old \citep[e.g.][]{portail2017,baba2020}.

Our view of the bar is only a snapshot in cosmic time, and so detailed studies of its structure and kinematics are important. Any significant departures from axial dynamics could indicate of strong perturbations and recent activity. Even in the presence of a high degree of bi-symmetry, a bar's subsequent evolution depends in part on its radial density profile \citep{sellwood2022}. While there has been substantial progress on the near side of the Galactic bar \citep{wegg2015,GaiaCollaboration2022Gaia}, the far side is largely unexplored. This is mostly due to the lack of reliable stellar distances beyond the reach of Gaia, which provides precise proper motions and radial velocities to stars beyond the Galactic center. However, distances with uncertainties $\lesssim$\,20\,\% are limited to within $\approx$\,3\,kpc from the Sun.  

\rev{To improve the precision of distance estimates, kinematics from Gaia can be combined with distances measured from period-luminosity (PL) relations of pulsating stars. Cepheid and RR Lyrae variables famously obey such a relation that scales with their dominant period of variability \citep{Leavitt1912Periods}, from which it is possible to estimate a star's absolute magnitude, and thus distance, knowing only its apparent magnitude and period. This has been used with great success for Cepheid and RR Lyrae stars to map the structure and kinematics of the Milky Way (e.g., \citealt{Chen2019Intuitive, Prudil2022Milky, Ripepi2017VMC}). However, these distances are subject to a variety of potential systematic errors, including inaccurate photometric measurements and metallicity biases \citep{Skowron2019Threedimensional}, and the presence of outliers \citep{Freedman2001Final}.}

\rev{A significantly more numerous class of pulsating stars that also obey PL relations are evolved red giants, commonly referred to as long-period variables (LPVs). LPVs are generally divided into semi-regular variables (SRVs) and Mira variables, based on the regularity and amplitude of their light curves \citep{catelanPulsatingStars2015}.} The periods observed in LPVs follow a series of distinct, parallel PL sequences that correspond to different pulsation modes. Their study has been greatly accelerated by ground-based large-scale variability surveys, such as MACHO \citep{Wood1999MACHO}, OGLE \citep{Soszynski2004Optical, Soszynski2009Optical}, ASAS-SN \citep{Auge2020Gaia}, as well as space-based missions like Hipparcos \citep{Bedding1998Hipparcos, Tabur2009Longterm, Tabur2010Periodluminosity}, CoRoT \citep{FerreiraLopes2015Variability}, and Kepler \citep{Banyai2013Variability}. A major breakthrough from Kepler was the realization that the PL relations in SRVs are a simple extension of the pattern of radial and non-radial solar-like oscillations observed in lower-luminosity red giants \citep{Stello2014Nonradial}, which follow a precise scaling between oscillations periods and amplitudes \citep{Huber2011Solarlike, Banyai2013Variability, Mosser2013PeriodLuminosity, Yu2020Asteroseismology}.

Several previous studies have used SRVs and Miras to probe the structure of the Milky Way \citep{Trabucchi2017New, Trabucchi2019Characterisation, Trabucchi2021Semiregular,Iwanek2022Threedimensional}, and demonstrated that calibrated PL sequences can yield distances to within 15\% precision \citep{Tabur2010Periodluminosity, Auge2020Gaia}. However, sample sizes have so far remained limited and did not include pulsation amplitudes as an additional constraint \citep[which increases the distance precision further;][]{Rau2019Calibrating}. In this paper, we present a new distance measurement method based on the period-amplitude-luminosity relations for SRVs to measure distances to \nstars\ stars observed by OGLE in the Galactic bulge. We combine this sample with proper motions and line-of-sight velocities from Gaia to perform the first investigation of the kinematic properties of the far side of the Galactic bulge.

\section{Methodology} \label{sec:method}
\subsection{Data} \label{sec:data}

\begin{figure}
    \centering
    \includegraphics[width=\linewidth]{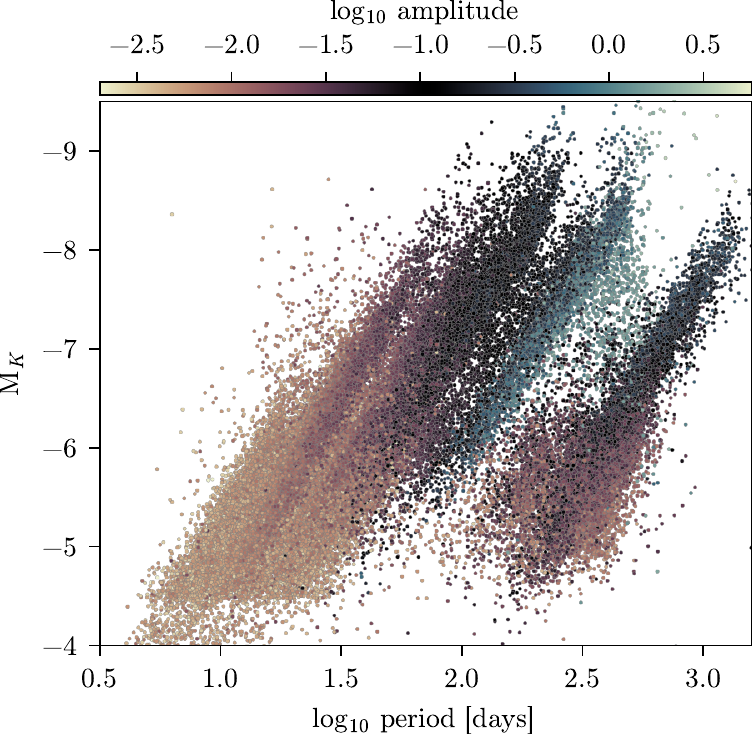}
    \caption{The Period-Amplitude-Luminosity relation of OGLE-III LPVs in the Large Magellanic Cloud. For clarity, we show only the dominant period (P1). The distinct sequences are due to different pulsation modes.}
    \label{fig:pl}
\end{figure}

\begin{figure*}
    \centering
    \includegraphics[width=0.8\linewidth]{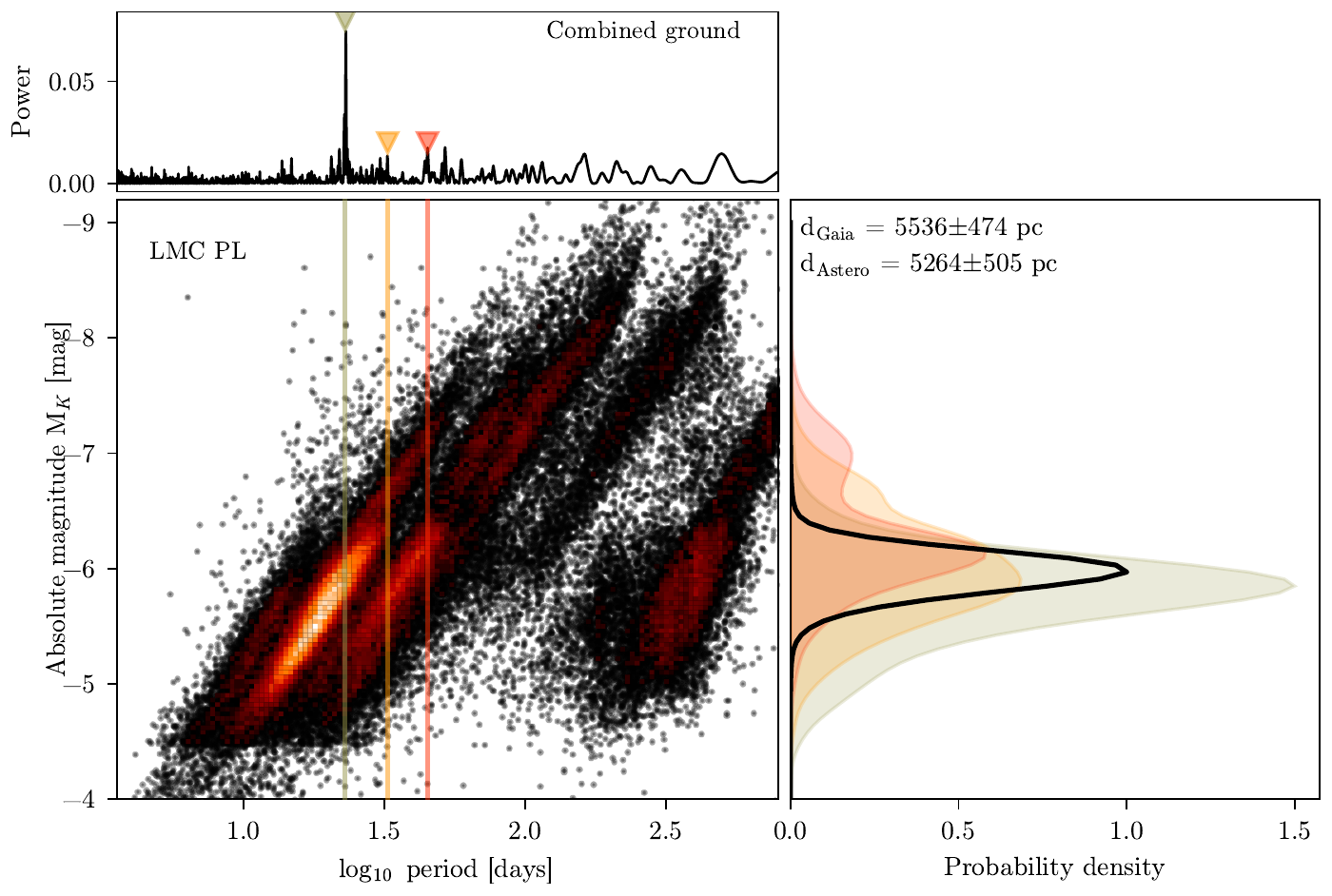}
    \caption{Distance estimation using the PL relation of the LMC for the semi-regular variable star KIC~7266343. \textbf{a)} The amplitude spectrum using ASAS-SN \citep{Shappee2014, Kochanek2017AllSky} photometry, shows three distinct peaks. \textbf{b)} The PL relation for the LMC colored by density. Slicing the peaks through the KDE representation of the PL relation gives \textbf{c)}, the normalized probability density of the absolute magnitude for each observed period. The black histogram depicts the combined PDF of each period.}
    \label{fig:pl_example}
\end{figure*}

We used data from two OGLE-III catalogs of long-period variables: in the Galactic bulge (\nstars\ stars; \citealt{Soszynski2013Optical}) and in the Large Magellanic Cloud (91,965 stars; \citealt{Soszynski2009Optical}). The catalogs include the three most prominent periods of variability, which were obtained through an iterative fitting and subtraction process. While the primary period was visually confirmed, the secondary and tertiary periods were not. To improve the accuracy, we screened the sample for signals that may be related to the synodic and sidereal lunar months and yearly aliases and removed any signals that were integer or half-integer multiples of these periods, within a tolerance of 0.1 days.

We included all variable types listed in the catalogs in our analysis. For both catalogs, we perform a positional (1\arcsec) cross-match with the 2MASS point source catalog (2MASS; \citet{Cutri20032MASS}) to obtain $J$, $H$, and $K$ band photometry. 

\subsection{The Period-Luminosity-Amplitude Relation} \label{sec:pl}

\begin{figure}
    \centering
    \includegraphics[width=\linewidth]{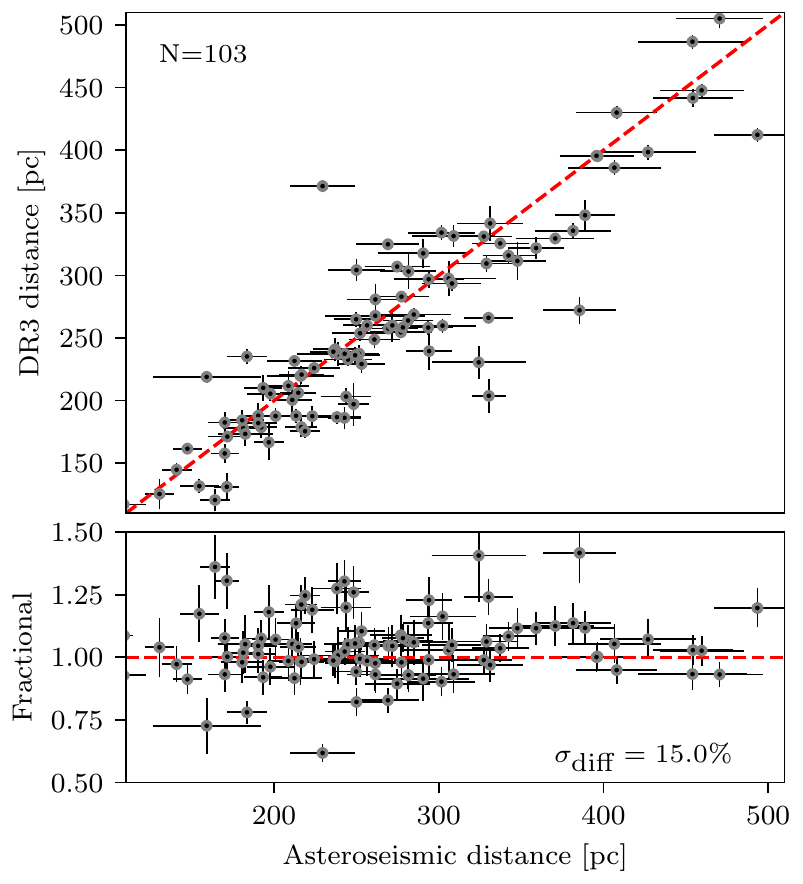}
    \caption{Comparison of distances from Gaia and PL \rev{(asteroseismic)} distances measured using our method for 103 nearby bright LPVs from \citet{Tabur2009Longterm}. The red line shows the 1:1 relation. The fractional standard deviation in the residuals is 15\%.}
    \label{fig:tabur_dist}
\end{figure}

\begin{figure*}[t]
    \centering
    \includegraphics[]{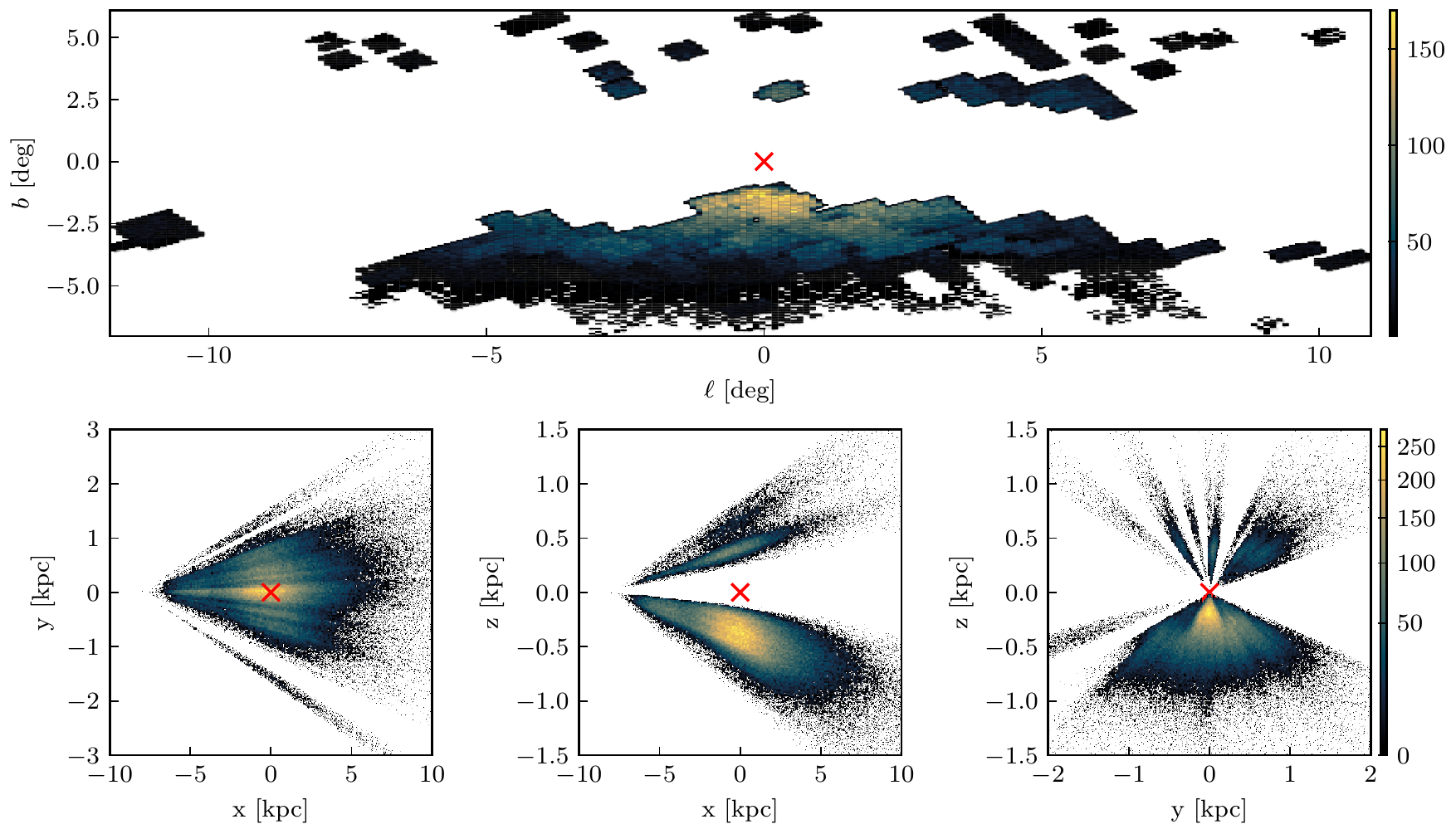}
    \caption{\rev{Upper panel: The Galactic bulge sample observed by OGLE in Galactocentric coordinates. Lower panels: Density of measured distances for the full sample of OGLE LPVs in Galactocentric Cartesian coordinates, with the Sun at ($-$8.122, 0) kpc. The stripes are due to the OGLE’s observational strategy, which did avoided the Galactic center.}}
    \label{fig:all_dist}
\end{figure*}

Figure \ref{fig:pl} shows the period-luminosity (PL) diagram of the Large Magellanic Cloud sample in the ($M_K$, $\log{P}$) plane, using a distance modulus of $\mu = 18.476\pm0.002$ mag and no correction for extinction \citep{Pietrzynski2019Distancea}. The parallel sequences correspond to different pulsation modes of different spherical degree and radial order  \citep[e.g.][]{Wood1999MACHO, Yu2021Asteroseismology}. The PL relations show a strong correlation with pulsation amplitude, with more luminous and longer period stars having higher amplitudes than their more rapidly variable counterparts, as predicted by \citet{Stello2007Multisite}. This effect is well-known and is useful to distinguish Mira variables from SRVs.

There is a natural spread in the PL relation of individual sequences. \citet{Lah2005Red} assumed that the spread is mainly due to distance variations of the stars within the LMC. From this, they derived a 3D map of the LMC by moving all points to their common sequences. If this assumption is correct, distances anchored on the LMC will have an intrinsic uncertainty floor set by this 3D structure. However, the depth of the LMC is expected to be less than 5\% of the total distance to the LMC itself \citep{Lah2005Red}.

The existence of multiple ridges in the PL diagram complicates the estimation of absolute magnitudes. A single observed pulsation period may intersect multiple ridges, making the absolute magnitude, and hence distance, uncertain. To address this issue, \citet{Tabur2010Periodluminosity} estimated absolute magnitudes by computing the probability density function (PDF) of the magnitude for multiple pulsation periods in a vertical strip within a $\pm2.5$\% of the period of interest. The resulting distributions were then used to estimate the probability of a star with a given period having a particular absolute magnitude. Multiplying these PDFs for the three strongest periods removes the degeneracy arising from multiple ridges, providing a reliable estimate of the magnitude.

In this work, we use a similar method but also take into account the amplitudes of the pulsations, and model the entire period-amplitude-luminosity space using a kernel density estimate (KDE). This method ensures that the resulting PDF is smooth and continuous, regardless of gaps in the data, and makes it simple to marginalize over the KDE. Moreover, spurious frequencies and amplitudes that do not conform with the PL relation will result in near-zero probability.

\rev{To construct the KDE, a bandwidth value is required to specify the width of the kernel, which dictates the smoothness of the estimated probability density function from the data. This bandwidth controls the `spread': a larger bandwidth results in a smoother estimate, whereas a smaller bandwidth results in a more peaked distribution because each data-point has a more localized effect. The optimal bandwidth thus best reproduces the underlying probability distribution of the data, without capturing small-scale noise or smoothing over important data features. The typical process for finding an optimal bandwidth involves splitting the data set into training and test samples, and then fitting the KDE on the training data with multiple bandwidths to compare against the test data. This is repeated until the predicted values most accurately reflect the test data to prevent over-fitting. However, due to the spread in the LMC PL relation, this approach results in a value that better represents the internal distribution of the LMC, which produces the characteristic spread around individual sequences.}

\rev{To overcome this limitation, we instead use the OGLE LPV Small Magellanic Cloud as a test sample \citep{Soszynski2011Optical}. We use a grid-search $k$-fold cross-validation method implemented in \textsc{scikit-learn} \citep{scikit-learn} to calculate the optimal bandwidth across a range of possible values. The `$k$-folds' refers to the method where the dataset is divided into $k$ equal-sized subsets. The training and evaluation process is repeated $k$ times, with each fold taking turns as the validation set while the remaining folds are used for training. This allows every data point to be used for validation exactly once, providing a more robust estimate of the model's performance compared to a single train-test split, where the average performance across all folds is used to select the best bandwidth. We find the resulting optimal bandwidth to be 0.26, with units corresponding to each dimension in the KDE (period (log$_{10}$ days), amplitude (log$_{10}$ ppt), absolute magnitude (mag)). We note that the bandwidth parameter in the KDE is functionally similar to the vertical strip width used by \citet{Tabur2010Periodluminosity}.}

To determine the absolute magnitude of a star, we calculate the probability density function for each measured pulsation period from the kernel density estimate and multiply them together. Figure~\ref{fig:pl_example} illustrates this method using a known Kepler LPV from \citet{Yu2021Asteroseismology} with three measured periods. We obtain a final estimate of the absolute magnitude in the chosen photometric band (in this case, the $K$-band) by multiplying the probability densities at each period. This distribution of absolute magnitudes is then converted to a distance modulus $\mu$ with the star's apparent magnitude $m$:
\begin{equation}
    \mu=m-M.
\end{equation}
Extinction is calculated iteratively from the \textsc{combined19} map of \citet{Bovy2015Galpy}. We first make an initial guess of the distance with no extinction, then iteratively calculate extinction until the value has converged. Applying this to our example star yields a distance of $d=5264 \pm 505$~pc, in excellent agreement with the Gaia DR3 distance of $d=5536\pm474$~pc \citep{Bailer-Jones2021Estimating}.

In general, the resulting PDF of a single star follows a normal distribution. To ensure the method has converged, we count the number of peaks in the distribution. Poorly constrained distances will have multiple peaks and are discarded for the remainder of this analysis. The distance modulus and its uncertainty are then taken as the median and standard deviation of the distribution. We note that the width of the final PDF depends on the bandwidth parameter of the KDE. Higher values lead to broader distributions and larger standard deviations. To avoid underestimating uncertainties, we ensure that each measured distance has at least an uncertainty given by the width of an individual sequence in the PL relation.

\subsection{Distance Validation}

We validated our method using the catalog of nearby, bright LPVs from \citet{Tabur2009Longterm}, which contains pulsation periods for 261 stars with precise astrometric distances from Gaia. \rev{We cross-matched this sample with the Gaia DR3 photogeometric distances \citep{Bailer-Jones2021Estimating}, and applied our method to measure PL distances using the pulsation periods from \citet{Tabur2009Longterm}.} The results are compared in Figure~\ref{fig:tabur_dist}. We remove stars that fail the distance estimation criterion as described in Sec.~\ref{sec:pl}, and ensure that the measured fractional distance uncertainty does not exceed 15\%, leaving us with 103 stars from the sample. We find that our PL distances agree well with the astrometric distances from \citet{Bailer-Jones2021Estimating}, with a residual offset of 4.1$\pm$1.5\% and a residual scatter of 15\%. This implies that random uncertainties dominate over systematic errors, and that the latter likely does not exceed the 1--2\% level, consistent with the systematic errors on the distance of the LMC. Note that the \citet{Tabur2009Longterm} catalog only reports periods and not amplitudes. We therefore expect the residual scatter to be an upper limit on the intrinsic uncertainty of our distance measurement method.

\section{A Period-Luminosity Distance to the Galactic Center} \label{sec:res}

\subsection{Distance Catalog}

\begin{table}
\centering
\begin{tabular}{lccr}
\hline
 ID & RA & Decl & Modulus [mag] \\
\hline
1 & 17:05:08.01 & -32:36:21.3 & 14.4$\pm$0.3 \\
2 & 17:05:11.04 & -32:52:33.7 & 15.0$\pm$0.3 \\
3 & 17:05:12.23 & -32:31:55.5 & 16.1$\pm$0.3 \\
4 & 17:05:14.81 & -32:53:17.5 & 15.2$\pm$0.2 \\
5 & 17:05:14.90 & -32:31:51.5 & 12.9$\pm$0.4 \\
6 & 17:05:15.11 & -32:58:44.8 & 14.5$\pm$0.2 \\
7 & 17:05:18.98 & -32:58:54.6 & 15.0$\pm$0.3 \\
8 & 17:05:20.57 & -32:31:29.3 & 14.9$\pm$0.2 \\
9 & 17:05:28.47 & -32:44:22.3 & 14.0$\pm$0.4 \\
10 & 17:05:29.17 & -32:34:39.5 & 15.2$\pm$0.2 \\
\hline
\end{tabular}
    \label{tab:distances}
    \caption{Results of the entire OGLE bulge sample. The ID column corresponds to the OGLE-BLG-LPV ID in the original catalog. The full table with additional columns is available online in electronic format.}
\end{table}

We applied the method described in Section~\ref{sec:method} to the full sample of \nstars\ LPVs observed by OGLE in the Galactic bulge. We used the three most significant peaks in the amplitude spectrum and their associated uncertainties according to the OGLE catalog. Of this sample, \ngood\ stars have a single peak in their absolute magnitude distribution, which we infer as having converged to a solution. The remainder of the stars has a multi-modal magnitude distribution from which no distance can be reliably estimated. The median distance uncertainty for the `good' sample is 12\%, with the lowest uncertainty being 7\%. For the rest of the paper, we only consider the 170,451 stars with less than 15\% distance uncertainties. 

To explore the distribution of this sample across the Milky Way, we converted the equatorial coordinates and distance into Galactic coordinates using \textsc{astropy} \citep{price-whelanAstropyProjectBuilding2018}. We use a Monte-Carlo method to determine uncertainties for the derived positions of our sample. For each star, we generate 100,000 samples of position and distance, drawn from their respective distributions. The Cartesian positions are then taken as the median and standard deviation. The uncertainty on the individual $J$ and $K$ band magnitudes from 2MASS are added in quadrature. Table~\ref{tab:distances} lists the derived distances and galactic coordinates for the full sample.

Figure~\ref{fig:all_dist} shows the distribution of these distances across the sky in Galactic Cartesian coordinates, where the Sun is at (0,0)~kpc. The density of measured distances is an indicator of the OGLE selection function, where the Galactic plane was not directly observed due to extinction. As expected, the density of stars peaks close to the center.

The top panel of Figure \ref{fig:gal_center} shows the fractional distance uncertainties of our sample. The median statistical uncertainty is $\approx$11\%, with a typical range of \mbox{$\approx$9--15\%}. Unlike astrometric distances, PL distances are largely independent of distance since pulsation periods and amplitudes, to first order, only depend on stellar luminosity. \rev{This enables the measurement of distances with uncertainties of $\approx$11\% out to 20\,kpc, highlighting the complementary nature of astrometry and asteroseismology for probing stellar distance in the Galaxy \citep{Huber2017Asteroseismology}.}

\subsection{Galactic Center Distance} \label{sec:dist}

\begin{figure}[t]
    \centering
    \includegraphics[width=\linewidth]{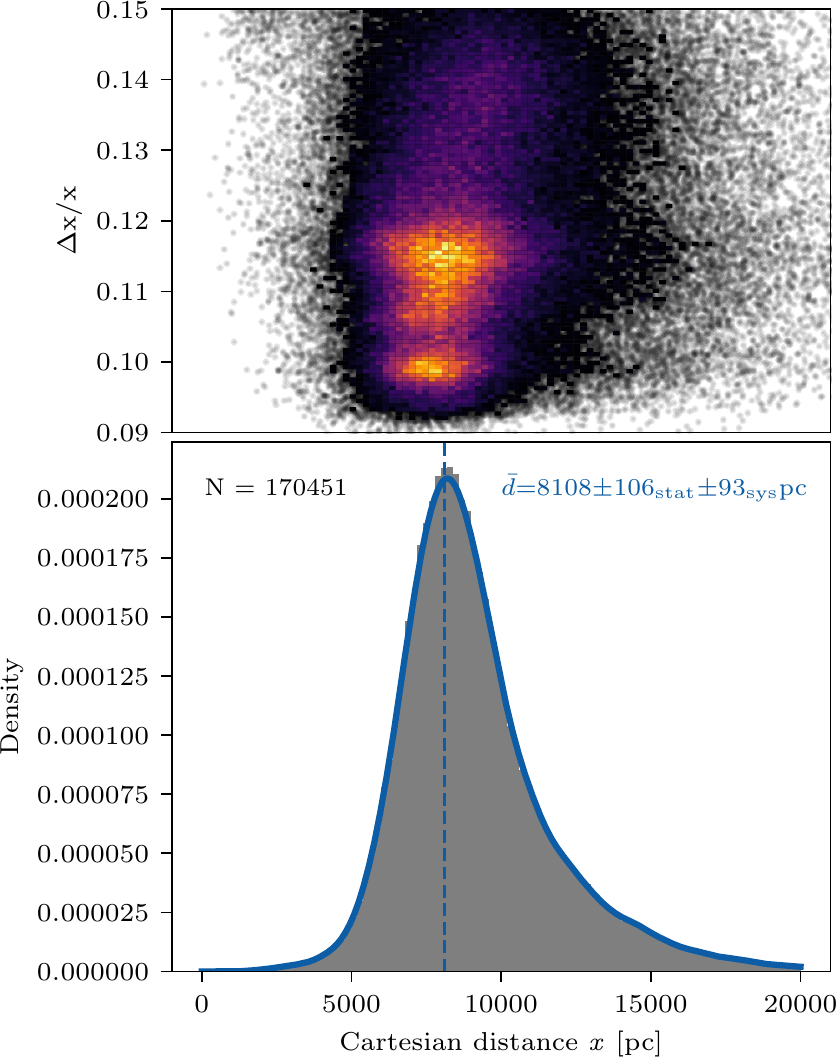}
    \caption{Top panel: Fractional distance uncertainty of the full sample as a function of distance. Bottom panel: Histogram of Galactic $x$ distances for \ngood\ stars with distance uncertainties less than 15\%. The blue dashed line shows the peak of the distribution measured from bootstrapping.}
    \label{fig:gal_center}
\end{figure}

The distance, $R_0$, to the Galactic center (GC) is of considerable interest as the anchoring distance for many cosmological models \citep{Bland-Hawthorn2016Galaxy}. Previous methods to measure the distance to the GC include can usually be divided into either direct or indirect measurements. The primary direct measurements involve either the parallax of Sgr B \citep{Reid2009TRIGONOMETRIC} or astrometric monitoring of orbits around Sgr A* \citep{Abuter2019Geometric}. A number of studies have also performed indirect measurements on the GC using variable stars, most commonly using Cepheids \citep{Chen2018Extremely, Griv2021Sun}, red clump giants \citep{Francis2014Two}, or Mira variables \citep{Matsunaga2009Nearinfrared}.

\rev{The large number of distances in our catalog enables a new indirect PL measurement of the distance to the GC. We can perform this measurement by assuming that the mass density profile of stars within the Milky Way is exponential \citep{Eilers2019Circular}. Indeed, simulations of stellar populations around the Galactic bulge imply the number density of red giant stars peaks at the Galactic center (see for example, figure 5 of \citealp{Clarke2019Milky}). With this assumption, we can estimate $R_0$ by measuring the mode of the Heliocentric Cartesian $x$ distances for our sample limited to stars where the fractional distance uncertainty is less than 15\%, and using a bootstrapping technique to estimate uncertainty, as shown in the bottom panel of Figure~\ref{fig:gal_center}. We perform 10,000 bootstrapping steps, where the histogram bins are randomly chosen at each step. The peak of the histogram is taken as the mode for each step, with the uncertainty given by the standard deviation of the final bootstrapped sample.}

We find a value for $R_0$ of \rnought~pc, which is in excellent agreement with the distance obtained by direct \rev{kinematic monitoring of stars orbiting} Sgr A* ($8178\pm13_{\rm stat}\pm22_{\rm sys}$~pc; \citealt{Abuter2019Geometric}) and the kinematic distance ($8123 \pm 120$~pc; \citealt{Leung2022Direct}). For our measurement, the statistical uncertainty arises from the bootstrapping of the histogram peak while the systematic uncertainty is dominated by the uncertainty in the LMC distance modulus \citep{Pietrzynski2019Distancea}.

\section{Kinematics of the Galactic bulge} \label{sec:kinematics}

\begin{figure}
    \centering
    \includegraphics[width=\linewidth]{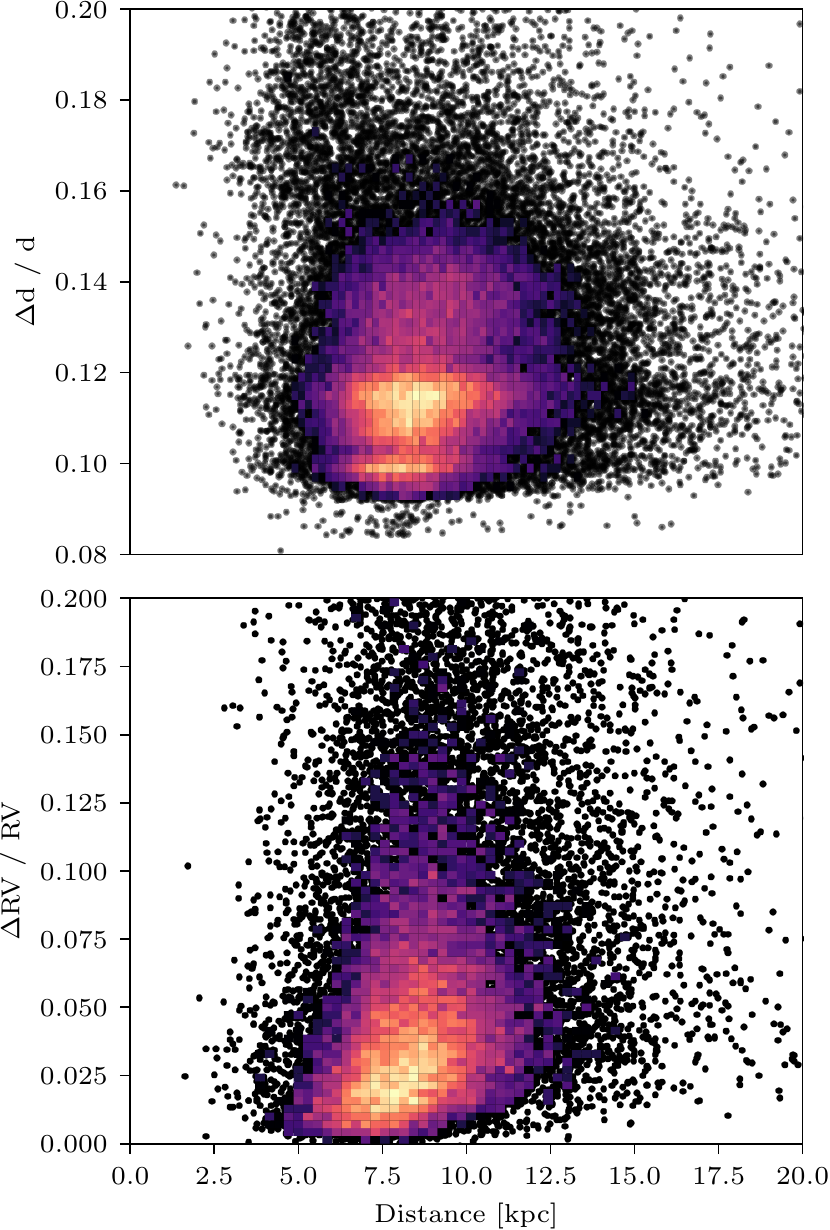}
    \caption{Uncertainties of the kinematic sample as a function of distance.}
    \label{fig:kinematic_uncertainties}
\end{figure}

\subsection{Kinematics Catalog}

\begin{figure*}
    \centering
    \includegraphics[width=\linewidth]{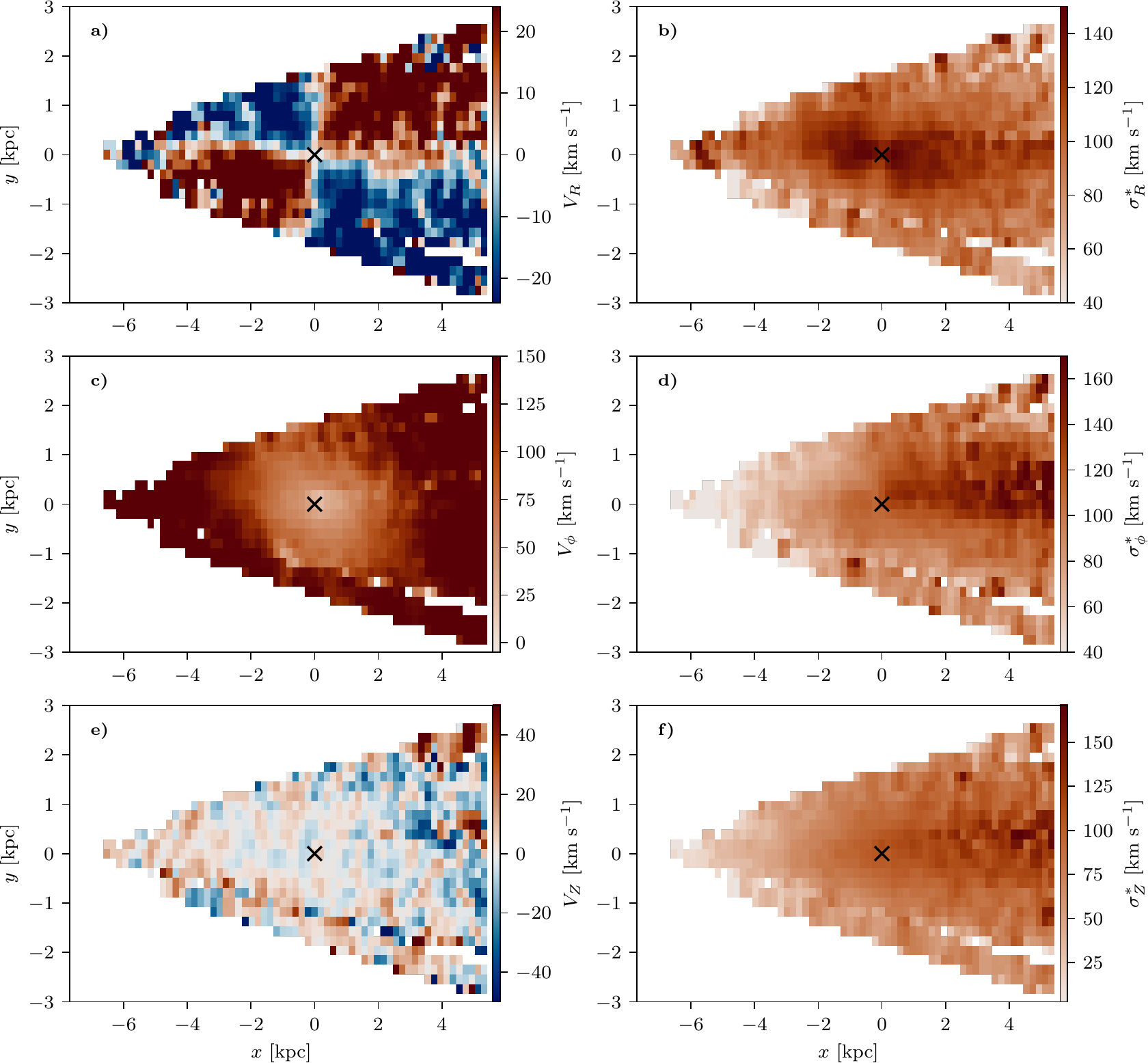}
    \caption{Velocity maps of the kinematic sample. The left-hand panels (a, c, e) show the mean radial ($V_R$), azimuthal ($V_\phi$), and vertical ($V_Z$) velocities respectively, while the right-hand panels (b, d, f) show their associated velocity dispersions. The black `x' indicates the Galactic center. Nearest neighbor smoothing has been applied to the data. The color ranges have been adjusted such that negative velocities are blue, and positive are red.}
    \label{fig:kinematics} 
\end{figure*}

To determine the 3D kinematics of our stars, we cross-matched our sample with the Gaia DR3 catalog \citep{Collaboration2021Gaia} using an angular matching radius of 1 arcsec and a magnitude match of $\pm$5 mag using OGLE I-band magnitudes. We adopted Gaia proper motions, radial velocities, and their associated uncertainties. Although the majority of stars in our sample have imprecise Gaia parallaxes, line-of-sight velocities are available for most stars brighter than $G\lesssim16$ mag \citep{Sartoretti2022Gaia}.

We again used the SkyCoord package implemented in \textsc{astropy} to convert our PL distances and Gaia positions into Galactocentric cylindrical coordinates. We used a right-handed coordinate system, where the Sun is at $-8.122$~kpc, and the solar motion is 12.9, 245.6, and 7.78~km s$^{-1}$ in the radial, rotational, and vertical directions, respectively, following v4.0 of \textsc{astropy}. Since there is uncertainty from both our measured distances and the Gaia radial velocities, we again follow a Monte-Carlo technique to derive uncertainties on the positions and velocities of the kinematic sample. For each star, we take 100,000 samples of our measured distances and uncertainty, as well as distributions of the proper motions and radial velocities. The distribution of coordinates for an individual star is then transformed into the Galactocentric frame. The final measured values are the median and standard deviation of the distribution.

We additionally made several quality cuts: the fractional uncertainties of the distance, proper motions, and radial velocities must all not exceed 15\%. We also removed all stars with a Reduced Unit Weight Error (RUWE) greater than 1.3, which is a typical indicator of binarity \citep{Belokurov2020Unresolved, Fabricius2021Gaia}. The final sample consists of 39,566 stars (Table~\ref{tab:kinematics}).

\subsection{Velocity maps}

\begin{figure}
    \centering
    \includegraphics[width=\linewidth]{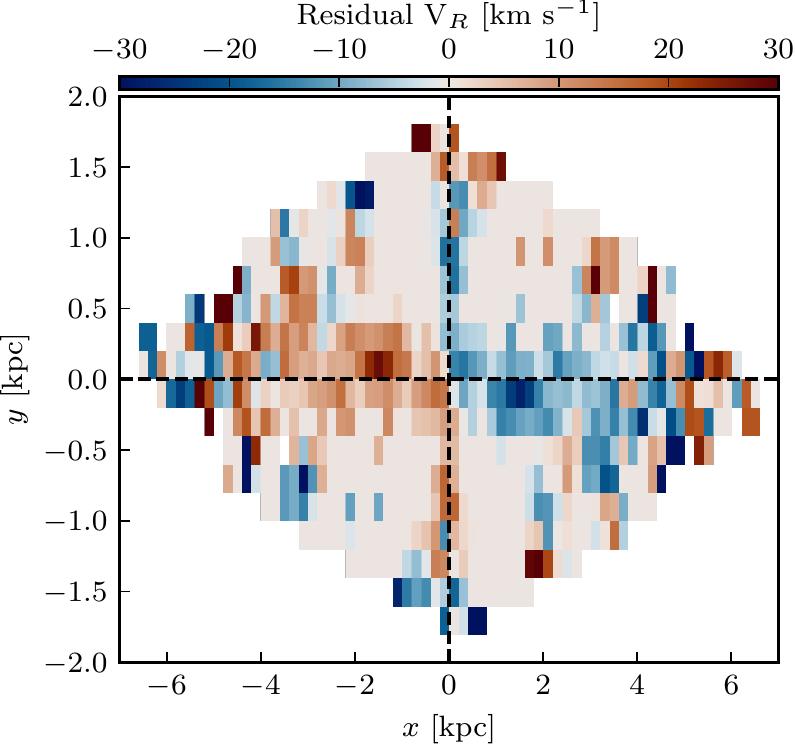}
    \caption{\rev{The residual radial velocity field, obtained by clipping and reflecting the velocities around the Galactic center, located at (0, 0) kpc.}}
    \label{fig:residual_vR}
\end{figure}

We use our kinematic sample to make maps of the Galactocentric rotational and radial velocities, as well as their associated velocity dispersion. To measure the kinematic properties of this sample, we followed the same approach and nomenclature as \citet{GaiaCollaboration2022Gaia}, which we briefly outline here. We subdivided the sample into 200 square parsec bins in the $x$-$y$ plane, where for each bin, we optimize for the velocity and its dispersion:
\begin{equation}
    \mathcal{L}(V_k, \sigma_k^*) = \frac{1}{2} \sum_j^N\Big[\ln(\sigma_k^{*^2} + \sigma_{v,k}^2) + \frac{(v_{k, j}) - V_k)^2}{\sigma_k^{*^2} + \sigma_{v,k}^{*^2}}\Big].
\end{equation}
Here, $V_k$ and $\sigma^*_k$ denote the mean and dispersion of the kinematic property $k$, covering each coordinate of the Galactocentric cylindrical frame ($R$, $\phi$, and $Z$), and $\sigma_{v,k}$ is the uncertainty on an individual measurement for that kinematic property. We then optimized each bin using the Nelder-Mead method implemented in \textsc{Scipy}. We discarded bins containing fewer than 5 stars.

\rev{The left panels of Figure~\ref{fig:kinematics} show the resulting velocity fields. The radial velocity map (Fig.~\ref{fig:kinematics}a) shows the expected bi-symmetric feature on both sides of the GC, with negative and positive values on either side of the major axis of the bar. This quadrupole feature is a characteristic of the mean streaming motion generated by the Galactic bar, and was identified by \citet{Bovy2019Life} and \citet{Queiroz2021Milky} using APOGEE line-of-sight velocities, and confirmed with Gaia DR3 \citep{GaiaCollaboration2022Gaia}. Our kinematic sample shows the bi-symmetric quadrupole on the opposite side of the center for the first time, confirming that it is reflected symmetrically. On both sides, there is a slight warping of the transition from negative to positive velocities.}

\rev{We demonstrate the degree of bi-symmetry in the quadrupole by a simple reflection of the radial velocity map around the Galactic center (Fig.~\ref{fig:residual_vR}). Since we only seek to show that the cloverleaf pattern is replicated, we clip the radial velocities to within $\pm20$ km$s^{-1}$ before reflecting.  The residuals show that the pattern is well replicated on either side of the GC, with a large discrepancy near the center. This is explained as a result of the slight warping effect seen in the original radial velocity map (Fig.~\ref{fig:kinematics}), and by the fact that the kinematic axis of the bar is not aligned with our line of sight.}

The azimuthal velocity (Fig.~\ref{fig:kinematics}c) is elongated along the major axis of the bar. Within the influence of the bar, the rotation is significantly slower along the bar axis, with another node at the GC. Interestingly, the vertical velocities (Fig.~\ref{fig:kinematics}e) show no discernible features within the range of the Galactic bulge, instead appearing corrugated. N-body simulations have suggested that Galactic bars tend to undergo one or multiple events of buckling instability, resulting in the formation of the boxy/peanut shape that should manifest as a distinct feature in kinematic space \citep{Athanassoula2002Morphology, Lokas2019Anatomy, Khoperskov2018Echo}. 

The flatness of our vertical velocities implies that the Galactic bulge is not currently undergoing a buckling episode, and is dynamically settled. This is in good agreement with three-dimensional density mapping with red-clump stars \citep{Wegg2013Mapping}. However, the selection function and observing strategy of the OGLE data means that we do not directly probe the Galactic center and inner disc.

We note that this effect is distinct from the formation of velocity sub-structures, or moving groups, as has been investigated by various authors, both near the Sun \citep{Trick2019,CraigChakrabarti2021}, and as a function of Galactocentric radius \citep{Lucchinimovinggroup}.  The formation of these moving groups \citep{CraigChakrabarti2021} can be heightened by dynamical interactions with dwarf galaxies \citep{ChakrabartiBlitz2009,Chakrabarti2019}.  Our focus in this paper is on the analysis of the vertical instability only, and we will investigate velocity sub-structures in a forthcoming paper.

\subsection{Velocity Dispersions}

The right panels of Figure~\ref{fig:kinematics} show the velocity dispersions for our sample.
The radial velocity dispersion map (Fig~\ref{fig:kinematics}b) shows a bi-symmetric feature that is aligned with the direction in which the radial velocity map changes sign. The GC is again the node of the quadrupole feature. This was also observed by \citet{GaiaCollaboration2022Gaia}, but the additional distances beyond the GC show that the pattern follows the expected dispersion predicted by N-body simulations of a barred, Milky-Way-like simulation \citep{Kawata2017Impacts}. The node of the radial velocity and dispersion is useful for measuring $R_0$ (see, e.g., \citealt{Leung2022Direct}).

The azimuthal and vertical velocity dispersions (Fig~\ref{fig:kinematics}d,f) show unexpected maxima $\sim 4$ kpc beyond the GC, possibly due to unaccounted-for uncertainties in the distances. Since all the velocity maps are produced using the same data, it is curious that only the azimuthal and vertical velocity dispersions show such a significant deviation. Previous work using Gaia DR3 notes a weak quadrupole feature in the dispersion, which we do not see here \citep{GaiaCollaboration2022Gaia}. However, the Gaia map is limited to a few hundred parsecs beyond the GC. We suggest two origins for the unusual features in the velocity dispersion. The most likely cause is increasing uncertainty with distance of line-of-sight velocity for stars lying beyond the GC. We show the fractional uncertainty in radial velocity and distance as a function of distance in Fig.~\ref{fig:kinematic_uncertainties}. While the distance uncertainty is mostly constant, the radial velocities tend to increase in uncertainty. We discuss these effects in more detail in Sec.~\ref{sec:comparison}. Another possibility is that the OGLE selection function and observing strategy (Fig.~\ref{sec:data}) only includes the top and bottom of the Galactic bulge. We may be seeing dispersions from stars at larger Galactic heights with different motions relative to the stars lying closer to the plane. We therefore recommend caution when interpreting results of dispersion maps beyond the Galactic center. 

\section{N-body/hydrodynamical simulations}
\label{sec:comparison}

\subsection{Simulation Details}

To investigate the effects of uncertainties on our results, we make use of a full N-body/hydrodynamical simulation of a barred Milky Way (MW) surrogate that includes a gaseous disk component. The simulation is approximated by a four-component system: a host dark matter (DM) halo ($M \approx 10^{12} ~M_\odot$), a `classic' stellar bulge ($M \approx 10^{10} ~M_\odot$), an 'old' stellar disk ($M \approx 4.5 \times 10^{12} ~M_\odot$), and a `cold' ($T \sim 10^3$~K) gas disk ($M \approx 4 \times 10^{9} ~M_\odot$), all of which are consistent with the Galaxy \citep{bla16a}. The setup and specific simulation details, as well as the evolution and properties of our MW surrogate without gas have been discussed at length by \citet{tep21v}. The model that includes a gaseous disk component has been presented and discussed in detail previously \citep{tep22x,dri23a}.

In brief, we generate initial conditions for all components using the self-consistent modeling module provided with the \agama\ library \citep{vas19a}. The gas disk is setup following the approach developed by \citet[][their `potential method']{wan10a}. The initial conditions are evolved for a total of $\sim5$ Gyr with the \ramses\ code \citep{tey02a}, which incorporates adaptive mesh refinement (AMR). During the run, a maximum spatial resolution of $\sim32$ pc is attained in the high-density regions (roughly within 10 kpc around the GC), and is progressively lower in general towards the galaxy edge.

The stellar disk mass and its scale length ($r_{\rm disk} \approx 2.5$ kpc) that we adopt for the initial conditions yield a disk-to-total mass ratio of $f_{\rm disk} \gtrsim 0.5$, implying that the disk is bar-unstable, with a bar formation timescale well below a Hubble time \citep{fuj18a,bla23}. In our model, a bar is formed within $\sim 2.2$ Gyr. The bar settles thereafter and evolves slowly over billions of years. Roughly 2.1 Gyr after the formation of the bar (or $\sim4.3$ Gyr since the start of the simulation), its structural and kinematic properties are in broad agreement with the corresponding properties of the Galactic bar \citep[cf.][]{bla16a}. In particular, its pattern speed is $\Omega_{bar} \approx$~43 km~s$^{-1}$~kpc$^{-1}$ and its length $\sim4.5$ kpc, which were estimated using the approach developed by \citet{deh23a}.  This model is similar to the long, slow bar model employed by \cite{DOnghia2020} in which a prominent velocity sub-structure known as Hercules is formed by the co-rotation resonance from stars orbiting the bar's Lagrange points \citep{Perez-Villegas2017Stellar, Lucchinietal2023}.

We thus focus our attention on this particular snapshot to study the impact of distance uncertainties on the kinematic maps of the stars within the bulge/bar region of the Galaxy. In this snapshot, the bar is aligned 26$^\circ$ to the line of sight, and the Sun is positioned at $(-8.122, 0)$ kpc in the ($x,y$) plane, following the convention of our kinematic sample.

\subsection{Influence of distance uncertainties}

\begin{figure*}[t]
    \centering
    \includegraphics[scale=1]{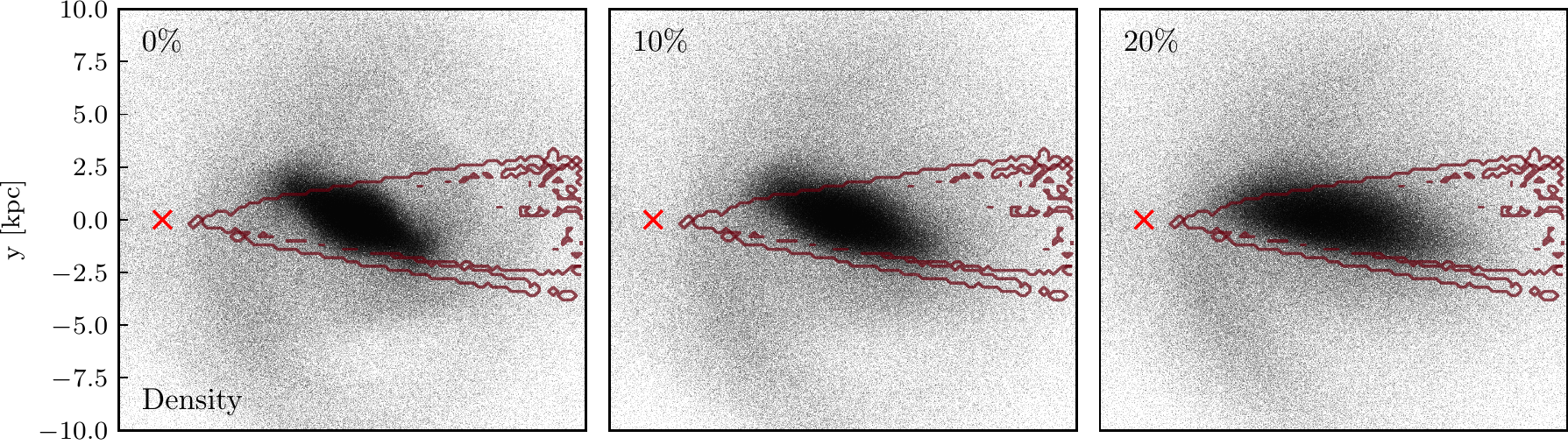}
    \includegraphics[scale=1]{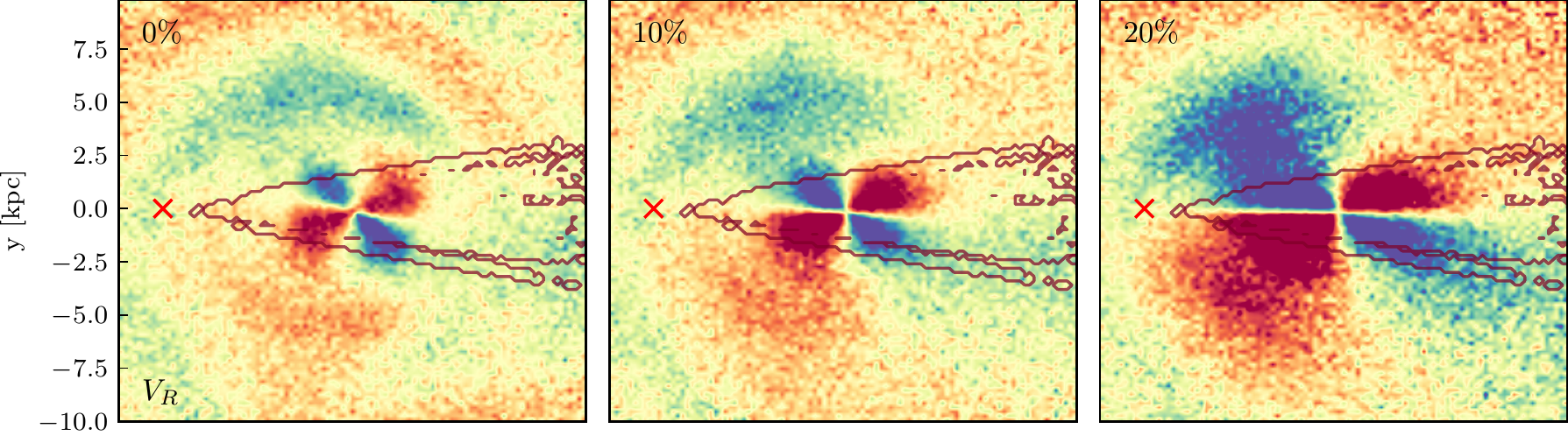}
    \includegraphics[scale=1]{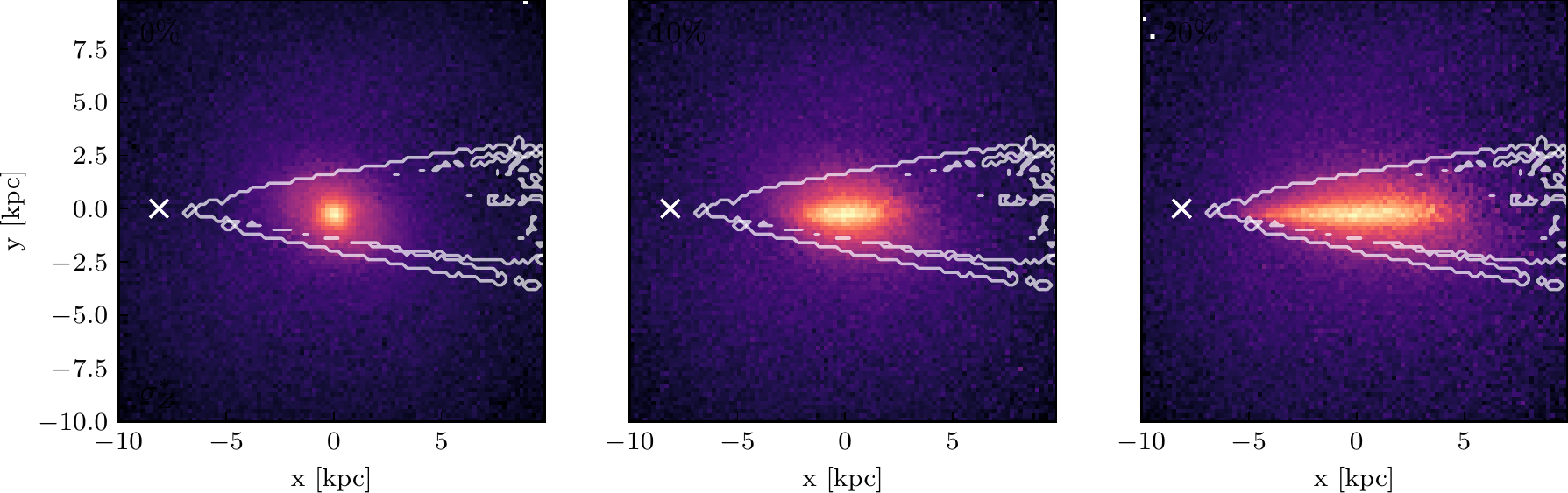}
    \caption{\rev{The impact of distance uncertainty on simulated data. Note that in these simulations, the Sun is situated at (-8.122, 0) kpc. In order from top to bottom, the panels represent the stellar density, radial velocity $V_R$, and vertical height dispersion $\sigma^*_{Z}$. The scatter in the distances increases from left to right (see text). The red contour outlines our sample observations.}}
    \label{fig:sims}
\end{figure*}
    
To assess the impact of distance uncertainties, we added a random scatter to the distances in the simulation. In one scenario, the scatter for each distance value was drawn from a normal distribution with a standard deviation of 10\%, and in a second scenario that value was 20\%. We selected stars within $\vert z\vert \leq 350$ pc of the Galactic plane. Since the simulation is initially centered at (0,0)~kpc, we transformed the coordinates into the Galactic frame and then converted to spherical coordinates. The radial components of the spherical coordinates were then sampled from a normal distribution with the chosen fractional uncertainty. Finally, the coordinates were transformed back to the Galactocentric Cartesian frame, with the new modified radial distance.  We do not correct for source crowding or variable opacity due to dust. We also do not account for the detailed selection function of the OGLE sample. However, neither of these effects are expected to change our main conclusions. 

Figure \ref{fig:sims} shows the effect of adding a random scatter to the distances in the simulation. We show three aspects of the simulation: density (top panels), radial velocity ($V_R$, middle panel), and $Z$ dispersion ($\sigma^*_Z$, bottom panel). In order from left to right, the scatter added to the distances was 0\%, 10\% and 20\%. We see that simulating distance uncertainties imprints a strong systematic bias, in particular on the quadrupole moment of the $V_R$ map. Importantly, the quadrupole effect fills a circle, not an ellipse, out to the bar termination radius ($\approx 5$ kpc). Also, the kinematic major axis (KMA) is aligned with the bar, unlike in the observed data (Fig.~\ref{fig:kinematics}), where the KMA is about half the bar angle to the line of sight. Furthermore, the unmodified velocity field shows a gentle azimuthal gradient between the blue-shifted and red-shifted quadrants, whereas in the observed data, the gradients are sharp. Based on Fig.~\ref{fig:sims}, we can explain both of these effects as arising from line-of-sight distance errors.

Additional evidence for this interpretation comes from the vertical velocity dispersion map $\sigma^*_Z$ (bottom panels of Figure \ref{fig:sims}). In the model, this map is highly circular and rises rapidly at a declining radius towards the GC, reflecting the deep Galactic potential. In the simulations, however, the observed $\sigma_z$ is stretched along the $y$ axis and has almost no azimuthal angle to the line of sight with increasing uncertainty. If distance errors dominate, this is to be expected because the inner bulge is spherical rather than elongated with the bar. 

The vertical height dispersion ($\sigma^*_Z$) with applied uncertainty does not show the same features as our data (Fig.~\ref{fig:kinematics}f). This strongly implies that our observed features in vertical height dispersion at large distances are instead due to the OGLE observing strategy.

In summary, we find that the inferred angle of the bar's kinematic major axis is significantly affected by the distance uncertainties, such that progressively increasing uncertainties lead to a smaller angle observed between the axis and the line of sight (Fig.~\ref{fig:sims}). Thus, by measuring the observed angle’s offset from the know bar position, we may be able to infer the distance uncertainties by degrading the simulation data with uncertainties at different levels and looking for a match.

\subsection{Comparison to data}

\begin{figure}
    \centering
    \includegraphics[width=\linewidth]{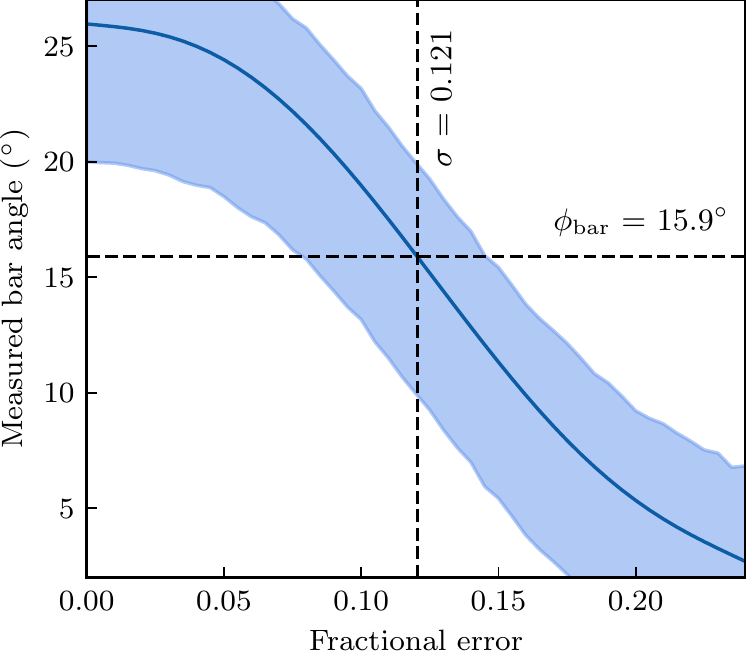}
    \caption{Measured kinematic bar angle as a function of distance uncertainty from simulations. The dark blue line is the angle measured with the simulation angle fixed at 26$^\circ$. The blue shaded region is obtained by shifting this angle between 20-30$^\circ$, in line with our uncertain knowledge of the true angle \citep{Bland-Hawthorn2016Galaxy}. The bar angle measured from our real data is marked as a horizontal black-dashed line, indicating a fractional distance uncertainty of approximately 12\%.}
    \label{fig:sims_rotate}
\end{figure}

We now provide a simple unbiased estimate of the distance uncertainties from our kinematic sample by comparing the bar angle measured from both the simulation and our data. To achieve this, we incrementally apply a fractional distance uncertainty, ranging from 0 to 20\% to the simulation data. At each step, we measure the Galactic bar angle from the simulation by fitting a simple 2D stretched and rotated Gaussian to the azimuthal $V_\phi$ map. While this is not a suitable model for most cases, it is adequate for our purposes.

We show the results in Figure~\ref{fig:sims_rotate}, indicating that increasing uncertainty results in a decreased bar angle measurement. This relation is non-linear. That is, the effect compounds rapidly above 10\% distance uncertainties. Curiously, this same result was noticed to a lesser extent in \citet{GaiaCollaboration2018Gaia}, where their measured bar angle was corrected by the estimated uncertainty. Applying the same analysis to our observed kinematic map of $V_\phi$ (Fig.~\ref{fig:kinematics}), we obtain a KMA angle of 15.9$^\circ$. According to the simulations with a fiducial bar angle of 26$^\circ$, this indicates a fractional distance error of approximately 12\%. Prior literature finds that the true value of the bar angle ranges from anywhere between 20 - 30$^\circ$ \citep{Bland-Hawthorn2016Galaxy}. \rev{Adjusting the simulation angle to these ranges implies the uncertainty can be anywhere from 8 to 15\%, with the nominal value of 12\% being in excellent agreement with our empirical uncertainty estimate of 11\% (Figure~\ref{fig:all_dist}).}

\rev{In comparison, Cepheid and RR Lyrae variables have distance uncertainties of approximately 1$-$5\% \citep{Owens2022Current, Chen2023Use}. While the SRV PL relation has a larger uncertainty, they are a significantly more abundant stellar tracer.}

\section{Conclusions} \label{sec:conc}

In this paper, we have used the period-amplitude-luminosity relation of evolved red giants (semi-regular variables) to determine distances to \nstars\ stars observed by OGLE in the Galactic bulge. Our main conclusions are as follows:
\begin{itemize}

    \item We demonstrate that the period-amplitude-luminosity relation of luminous red giants (semi-regular variables) can be used to measure distances with a random uncertainty of $\approx$11\% and systematic errors not exceeding $\approx$\,1--2\%. The method depends primarily on stellar luminosity and thus is capable of providing accurate distances out to several tens of kpc for hundreds of thousands of red giants.

    \item We measure a period-luminosity distance to the Galactic center of $R_0$ = \rnought\ pc. This value is consistent with \rev{kinematic} monitoring of stars orbiting Sgr A*, and the uncertainty is dominated by the distance uncertainty of the Large Magellanic Cloud, which is used to anchor our distance scale.

    \item By cross-matching our distance catalog with kinematic data from Gaia, we provide the first constraints on the Milky Way's velocity field beyond the Galactic center. We show that the $V_R$ quadrupole from the bar's near side is reflected with respect to the Galactic center, indicating that the bar is both bi-symmetric and aligned with the inner disk, and therefore dynamically settled along its full extent.

    \item The kinematic map  of vertical height, $V_Z$, has no major structure in the region of the Galactic bulge. This implies that the bulge is not currently undergoing a buckling episode and is dynamically settled, in good agreement with the three-dimensional density mapping of red-clump stars.

    \item We demonstrate with N-body simulations that distance uncertainty plays a major factor in the alignment of the major axis of the bar and the distribution of velocities. In particular, we show that distance uncertainties of around 20\% are sufficient to completely warp the major axis of the bar measured from kinematics. This implies that distance uncertainties must be taken into account when measuring detailed properties of the bar/bulge from kinematic data.

\end{itemize}

The results presented here provide the first glimpse of the potential for using the pulsations of luminous red giants to map the kinematic structure of the far reaches of the Milky Way. Applying our method to light curves from current and future all-sky ground-based surveys, such as ATLAS \citep{Tonry2018ATLAS}, ASAS-SN \citep{Shappee2014, Kochanek2017AllSky}, ZTF \citep{Bellm2014Zwicky} and LSST \citep{Ivezic2019LSST}, will allow similar kinematic maps out to a significant fraction of the Milky Way beyond the Galactic bulge. Combined with masses and ages from more detailed asteroseismic datasets of nearer stellar populations observed by Kepler/K2 and TESS \citep{Stello2014Nonradial, Stello2022TESS, Jackiewicz2021SolarLike}, these datasets will provide insight into the formation and evolution of our Milky Way.

\begin{acknowledgments}

    D.R.H., D.H., R.S., and SC acknowledge support from National Science Foundation (AST-2009828) and from the Research Corporation for Science Advancement through Scialog award $\#$26080. D.H.\ also acknowledges support from the Alfred P. Sloan Foundation and the Australian Research Council (FT200100871).
    T.T.G. acknowledges financial support from the Australian Research Council (ARC) through an Australian Laureate Fellowship awarded to J.B.H. We acknowledge the use of the National Computational Infrastructure (NCI),  supported by the Australian Government, and accessed through the NCI Adapter Scheme 2023 (Principal Investigator: T.T.G.; Chief Investigator: J.B.H.).
    T.R.B. acknowledges support from an Australian Research Council Laureate Fellowship (FL220100117).
    N.S.\ acknowledges support from the National Science Foundation through the Graduate Research Fellowship Program under Grant 1842402.

\end{acknowledgments}

\vspace{5mm}

\software{astropy \citep{astropycollaborationAstropyCommunityPython2013},
    numpy \citep{Oliphant2015Guide},
    matplotlib \citep{Hunter2007Matplotlib},
    Gala \citep{Price-Whelan2017Gala},
    scipy \citep{Virtanen2019SciPy},
    scikit-learn \citep{scikit-learn}
}

\appendix
\section{Table of kinematics}
\begin{table}
\centering
\begin{tabular}{rrrrrrr}
\hline
ID & x [pc] & y [pc] & z [pc] & V$_{\textrm R}$ [kms$^{-1}$] & V$_{\phi}$ [kms$^{-1}$]  & V$_{\textrm Z}$ [kms$^{-1}$] \\
\hline
2 & 1667.8 & -1397.7 & 856.5 & 86$\pm$37.8 & 29$\pm$56.4 & -1$\pm$2.2 \\
3 & 8270.4 & -2259.1 & 1478.4 & -33$\pm$3.2 & 206$\pm$62.2 & -75$\pm$11.5 \\
4 & 2833.3 & -1564.5 & 952.5 & 185$\pm$18.8 & 150$\pm$83.4 & -3$\pm$2.6 \\
8 & 1109.9 & -1268.4 & 838.5 & 50$\pm$27.2 & 8$\pm$33.6 & -25$\pm$4.1 \\
10 & 2503.0 & -1464.4 & 951.4 & -113$\pm$22.7 & 135$\pm$14.9 & -117$\pm$12.6 \\
11 & 4353.7 & -1738.2 & 1098.8 & -17$\pm$5.6 & 200$\pm$47.5 & 75$\pm$8.2 \\
12 & 2641.6 & -1552.3 & 911.9 & -115$\pm$15.9 & 88$\pm$14.7 & 43$\pm$5.2 \\
14 & 1387.7 & -1364.7 & 812.9 & -14$\pm$36.8 & 153$\pm$41.9 & 58$\pm$8.1 \\
17 & 6777.1 & -2052.2 & 1317.8 & 185$\pm$8.8 & 511$\pm$126.2 & -74$\pm$13.5 \\
20 & 11023.8 & -2723.1 & 1624.0 & 35$\pm$3.3 & 93$\pm$40.1 & 4$\pm$0.9 \\
\hline
\end{tabular}
    \label{tab:kinematics}
    \caption{Truncated results of the kinematic sample. The
ID column corresponds to the OGLE-BLG-LPV ID in the
original catalog.}
\end{table}

\bibliography{library}{}

\bibliographystyle{aasjournal}

\end{document}